\documentstyle[prd,aps,epsf]{revtex}
\draft
\begin{document}
%\newcommand{\gsim}
%   {\mathrel{\raise.3ex\hbox{$>$\kern-.75em\lower1ex\hbox{$\sim$}}}}
%\newcommand{\lsim}
%   {\mathrel{\raise.3ex\hbox{$<$\kern-.75em\lower1ex\hbox{$\sim$}}}}
%\newtheorem{th}{Theorem}
%\newtheorem{pr}{Proposition}
%\newtheorem{lm}{Lemma}

\thispagestyle{empty}
{\baselineskip0pt
\leftline{\baselineskip16pt\sl\vbox to0pt{\hbox{\it Department of Physics}
               \hbox{\it Kyoto University}\vss}}
\rightline{\baselineskip16pt\rm\vbox to20pt{\hbox{KUNS-1623}
%           \hbox{December 1998}
            \hbox{\today} 
\vss}}%
}
\vskip1cm
%\vfill
\begin{center}{\large \bf
Naked Singularity Explosion}
\end{center}
\begin{center}
 {\large 
Tomohiro Harada
\footnote{ Electronic address: harada@tap.scphys.kyoto-u.ac.jp}
, Hideo Iguchi 
\footnote{ Electronic address: iguchi@tap.scphys.kyoto-u.ac.jp}}\\
{\em Department of Physics,~Kyoto University,} \\
{\em Kyoto 606-8502,~Japan}\\
and\\ {\large Ken-ichi Nakao
\footnote{ Electronic address: knakao@sci.osaka-cu.ac.jp}}\\
{\em Department of Physics,~Osaka City University,} \\
{\em Osaka 558-8585,~Japan}\\
\end{center}
%\vfill

\begin{abstract}
It is known that the  
gravitational collapse of a dust ball
results in naked singularity formation
from an initial density profile which is physically reasonable.
In this paper, we show that explosive radiation is emitted 
during the formation process of
the naked singularity. 
 
\end{abstract}


It is known that the gravitational 
collapse of an inhomogeneous dust ball
results in shell-focusing naked singularity 
formation~\cite{es1979,christodoulou1984,newman1986,jd1993}.
It has been also shown that the naked singularity 
formation is possible from 
the spherical collapse of a perfect 
fluid with a very soft equation of state
~\cite{op1987,op1988,op1990,harada1998}.
Moreover, a kind of runaway collapse
in Newtonian gravity is similar to the naked singularity 
formation in general relativity in many respects.
These strongly suggest that the gravitational collapse 
will often involve drastic growth of spacetime curvature
outside the event horizon.

From this point of view, a number of researchers have examined 
emission during the naked singularity formation.
In classical theory,
Nakamura, Shibata and Nakao~\cite{nsn1993}
suggested that the forming naked singularity
in the collapse of a prolate spheroid
may be a strong source of gravitational waves.
Recently, Iguchi, Nakao and Harada~\cite{inh1998} and
Iguchi, Harada and Nakao~\cite{ihn1999,ihn2000} examined 
the behavior of nonspherical linear perturbations of the 
spherical dust collapse. They reported rather milder instability in
\cite{ihn2000}. 

Hawking~\cite{hawking1975} showed that
thermal radiation is emitted from the 
gravitational collapse to a black hole by quantum effects.
In the formation of a globally naked singularity,
the spacetime curvature grows unboundedly
and the strongly curved region can be seen by a distant observer,
unlike in the formation of a black hole.
This fact suggests that the forming 
naked singularity may be a strong source
of radiation owing to quantum effects.
In this context, Ford and Parker~\cite{fp1978} 
calculated the radiation during the formation of a 
shell-crossing naked singularity and found that
the luminosity is finite.
This will be because the shell-crossing singularity is 
very weak.
It is known that the shell-focusing singularity is stronger 
than the shell-crossing singularity.
Hiscock, Williams and Eardley~\cite{hwe1982}
showed the diverging luminosity during the formation of
the shell-focusing singularity in the 
spherically symmetric, self-similar collapse of
a null dust.
Barve, Singh, Vaz and Witten~\cite{bsvw1998a}
also showed the diverging luminosity
during the shell-focusing singularity 
in the self-similar collapse of a dust ball.
See also~\cite{bsvw1998b,vw1998}.

The last two examples in which the diverging luminosity is emitted 
are self-similar collapse.
However, the self-similar collapse is a particular solution 
among gravitational
collapse solutions in general relativity.
Moreover, it is uncertain whether or not the central part 
of the realistic spherical collapse with nonzero pressure 
tends to be self-similar in strong-gravity regime such as
shell-focusing singularity formation (cf. \cite{harada1998}).
For the self-similar collapse of a dust ball,
it has been shown that the redshift at the center 
diverges to infinity and
that the curvature strength of the naked singularity 
is very strong.
In fact, it is known that this solution 
does not allow an initial density profile which is a $C^{\infty}$
function with respect to the local Cartesian coordinates.
For $C^{\infty}$ case, the features are much different.
The redshift is finite and the curvature strength is not 
very strong~\cite{djd1999,hni1999,dwivedi1998}.
Though Einstein equation does not require
such strong differentiability of initial data,
we usually set such initial data in most 
astrophysical numerical simulations.
We should also comment that this model of the 
collapsing dust ball will be valid until perturbations sufficiently
grow due to the reported mild instability.
In this paper, we examine radiation 
during the naked singularity formation
in the collapse of an inhomogeneous dust ball 
from an initial density profile which is a $C^{\infty}$ function.
We use the units in which $G=c=\hbar=1$.

We consider both minimally and conformally coupled 
massless scalar fields in four dimensional spacetime
which is spherically symmetric and asymptotically flat.
Let $u$ and $v$ be null coordinates such that
they are written as $u\approx T-R$ and $v\approx T+R$
in the asymptotic region
with the quasi-Minkowskian spherical coordinates $(T,R,\theta,\phi)$.
An outgoing null ray $u=\mbox{const}$ arriving on ${\Im}^{+}$
can be traced back through the geometry becoming an incoming null ray 
$v=\mbox{const}$ originating from ${\Im}^{-}$ with $v$.
This gives the relation between $u$ and $v$ and we define
the function $G(u)$ by $v\equiv G(u)$.
Here, we assume that geometrical optics approximation is valid,
which implies that the trajectory of the null ray gives 
a surface of a constant phase of the scalar field.
Then, the luminosity $L_{lm}$ for the minimally coupled scalar field 
and the luminosity $\hat{L}_{lm}$ for the conformally coupled scalar field 
for fixed $l$ and $m$ are given 
through the point-splitting regularization
as~\cite{fp1978}
\begin{equation}	
L_{lm}
=\frac{1}{48\pi}\left(\frac{G^{\prime\prime}}{G^{\prime}}\right)^2
-\frac{1}{24\pi}\left(\frac{G^{\prime\prime}}
{G^{\prime}}\right)^{\prime},~
\hat{L}_{lm}=\frac{1}{48\pi}
\left(\frac{G^{\prime\prime}}{G^{\prime}}\right)^2.
\label{eq:llmhatllm}
\end{equation}	
This implies that the luminosity depends on
how the scalar field couples with gravity.
However, if $G^{\prime\prime}/G^{\prime}|_{u=a}=
G^{\prime\prime}/G^{\prime}|_{u=b}$ 
holds,
the amounts of the radiated energy 
during $a\le u\le b$ for the both fields are the same.
We should note that the 
geometrical optics approximation is only valid
for smaller $l$.
Hereafter we omit the suffix $l$ and $m$.
It is noted that these results are free of 
ambiguity coming from local curvature
because the regularization is done
in flat spacetime.

The spherically symmetric collapse of a dust fluid is
exactly solved~\cite{tolman1934,bondi1947}.
The solution is called the Lema\^{\i}tre-Tolman-Bondi (LTB)
solution.
For simplicity, we assume the marginally bound collapse.
The metric is given by
\begin{equation}
ds^2=-dt^2+R_{,r}^{2}(t,r)dr^2
+R^2(t,r)d\Omega^{2}
\label{eq:metric}
\end{equation}
in the synchronous comoving coordinates with
$d\Omega^{2}\equiv d\theta^2+\sin^{2}\theta d\phi^2$.
The energy density is given by
\begin{equation}
\epsilon =\frac{F^{\prime}(r)}{8\pi R^2 R_{,r}},
\label{eq:epsilon}
\end{equation}
where $F(r)$ is equal to twice the 
Misner-Sharp mass.
We rescale the radial coordinate $r$ as $R(0,r)=r$.
Then, $R$ is given as
\begin{equation}
R=r \left(1-\frac{3}{2}\sqrt{\frac{F}{r^3}} t\right)^{2/3}.
\label{eq:R}
\end{equation}
The singularity occurs at the time 
$ t_{s}(r)\equiv (2/3)\sqrt{r^3/F}$.
We denote the time of occurrence of singularity
at the center as $t_{0}\equiv t_{s}(0)$.
From Eq.~(\ref{eq:epsilon}), if we require 
that an initial density profile
at $t=0$
is a $C^{\infty}$ function with respect to the local Cartesian coordinates,
$F(r)$ is expanded around $r=0$ as
\begin{equation}
F(r)=F_{3}r^3+F_{5}r^5+F_{7}r^7+\cdots,
\label{eq:expansion}
\end{equation}
where we assume that $F_{3}$ is positive.
If we assume the marginally bound collapse 
and the mass function
as Eq.~(\ref{eq:expansion}),
then the central shell-focusing singularity is naked 
if and only if $F_{5}$ is 
negative~\cite{es1979,christodoulou1984,newman1986,jd1993,sj1996,jjs1996,jj1997}.
At an arbitrary radius $r=r_{sf}$,
the LTB spacetime can be matched with the Schwarzschild spacetime 
\begin{equation}
ds^2=-\left(1-\frac{2M}{R}\right)dT^2+\left(1-\frac{2M}{R}\right)^{-1}
dR^2+R^2d\Omega^{2},
\end{equation}
with $M=F(r_{sf})/2$.
The relation between $T$ and $t$ is given at $r=r_{sf}$ as
\begin{equation}
T=t-\frac{2r^{3/2}}{3\sqrt{2M}}-2\sqrt{2MR}+2M\ln\frac{\sqrt{R}+\sqrt{2M}}
{\sqrt{R}-\sqrt{2M}}.
\end{equation}
The null coordinates $u$ and $v$ in the Schwarzschild 
spacetime are defined as
\begin{equation}
u\equiv T-R_{*},~
v\equiv T+R_{*}
\label{eq:uv}
\end{equation}
with $R_{*}\equiv R+2M\ln[(R/2M)-1]$.

We can determine the function $G$ by solving the trajectories 
of outgoing and ingoing null rays in the dust and
calculating $u$ and $v$ through Eq.~(\ref{eq:uv})
at the time when the outgoing and ingoing null rays reach 
the surface boundary.
The trajectories of null rays in the dust are given by
the following ordinary differential equation
\begin{equation}
\frac{dt}{dr}=\pm R_{,r},
\label{eq:dtdr}
\end{equation} 
where the upper and lower signs denote
outgoing and ingoing null rays, respectively. 
We have numerically solved Eq.~(\ref{eq:dtdr})
by the Runge-Kutta method of the fourth order.
We have carried out the quadruple precision calculation
for retaining accuracy. 
We have chosen the mass function as
$	F(r)=F_{3}r^3+F_{5}r^5$.
We find that the central singularity is globally naked for very 
small $r_{sf}$ if we fix the 
value of $F_{3}$ and $F_{5}$. 
Although we have calculated several models,
we only display the numerical results for the model with
$F_{3}=1$, $F_{5}=-2$ and $r_{sf}=0.02$ in an arbitrary unit
because the features are the same if the singularity 
is globally naked.
The total gravitational mass $M$ is given by $M=3.9968\times 10^{-6}$
for this model. 

\begin{figure}
	\centerline{\epsfxsize 8cm \epsfysize 6cm \epsfbox{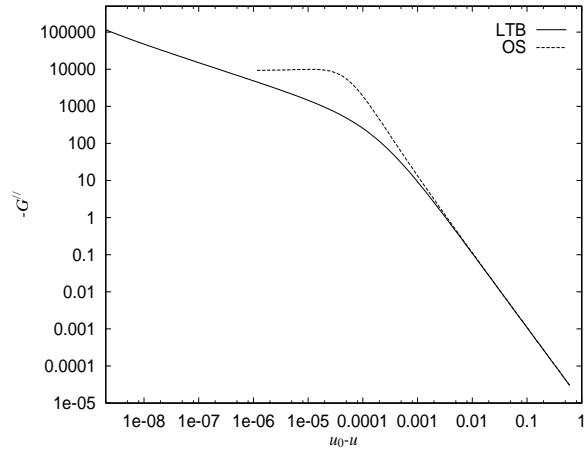}}
~(b)
	\caption{(a) $G^{\prime}(u)$ and (b)
 $G^{\prime\prime}(u)$ for the LTB spacetime and the OS spacetime.}
\label{fg:g_comb}
\end{figure}

See Fig.~\ref{fg:g_comb}. 
We define
$u_{0}$ as the retarded time of the earliest 
light ray which originates from the singularity.
In Fig.~\ref{fg:g_comb}(a), the first 
derivative $G^{\prime}(u)$ is plotted.
This shows that $G^{\prime}(u)$ does not diverge
but converge to some positive value $A$ with $0<A<1$.
In Fig.~\ref{fg:g_comb}(b), it is found that
the second derivative $G^{\prime\prime}(u)$ does diverge 
as $u\to u_{0}$.
This figure shows that the behaviors of growth of $G^{\prime\prime}(u)$ 
are different each other for
$10^{-4}\alt u_{0}-u $ and for
$0 < u_{0}-u \alt 10^{-4}$.
For $10^{-4}\alt u_{0}-u $ , the dependence on $u$ is written as
$	G^{\prime\prime}\propto -(-u)^{-2}$,
while, for $0 < u_{0}-u \alt 10^{-4}$, the dependence is written as
$	G^{\prime\prime}\propto -(u_{0}-u)^{-1/2}$.
Here we determine a dimensionful 
constant of proportion by physical consideration.
We assume that the early time behavior is due to
the collapse of the dust as a whole while the late time
behavior is due to the growth of the central curvature.
This assumption will be justified later.
First we should note that the coefficient must be
written using the initial data 
because it must not depend on time.
Next we should note that we can regard any $t=\mbox{const}<t_{0}$ 
hypersurface as an initial hypersurface.
Therefore, 
the coefficient must be independent of
the choice of an initial slice. 
For the early time behavior, the only possible quantity is 
the gravitational mass $M$ of the dust cloud.
For the late time behavior, 
the only possible quantity
is $\omega_{s}\equiv l_{0}^{6}/t_{0}^7 
= (3/2)^{7}F_{3}^{13/2}(-F_{5})^{-3}$, where $t_{0}$ is given as
$t_{0}=(2/3)F_{3}^{-1/2}$
and $l_{0}$ denotes the
scale of inhomogeneity defined as
$l_{0}\equiv(-F_{5}/F_{3})^{-1/2}$.
We call $\omega_{s}$ a singularity frequency. 
The $\omega_{s}$ is 
independent of the choice of an initial slice
because the mass function $F(r)$ is written in terms of $R$ as
\begin{equation}
F(r)=F_{3}\left(\frac{t_{0}-t}{t_{0}}\right)^{-2}R^3
+F_{5}\left(\frac{t_{0}-t}{t_{0}}\right)^{-13/3}R^5+\cdots,
\end{equation} 
around the center.
Thus, we can write 
$	G^{\prime\prime}\approx -f_{e}M(-u)^{-2}$ for $-u\gg M$, and
$	G^{\prime\prime}\approx -f_{l}A
	\omega_{s}^{1/2}(u_{0}-u)^{-1/2}$ for $0<u_{0}-u\ll \omega_{s}^{-1}$,
where $f_{e}$ and $f_{l}$ are dimensionless 
positive constants of order unity.
The turning point from the early time behavior 
to the late time behavior is roughly estimated as
$u_{0}-u\approx (M\omega_{s})^{2/3}\omega_{s}^{-1}$.
These estimates show a good agreement with the numerical results. 

\begin{figure}[htbp]
	\centerline{\epsfxsize 8cm \epsfysize 6cm \epsfbox{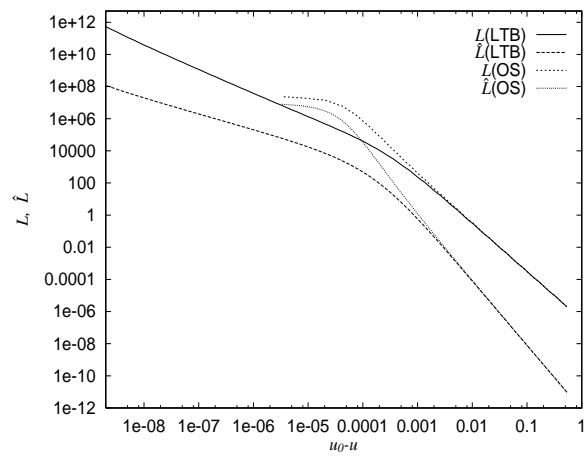}}
	\caption{Luminosity for minimally and conformally
	coupled scalar fields in the LTB spacetime and in the OS spacetime}
	\label{fg:p_comb}
\end{figure}

Let us consider the luminosity which is	calculated by 
Eq.~(\ref{eq:llmhatllm}).
The numerical results are displayed in Fig.~\ref{fg:p_comb}.
We can also write an analytic
expression for the luminosity 
using Eq.~(\ref{eq:llmhatllm}).
In Fig.~\ref{fg:g_comb}(a), we can find 
$	G^{\prime}\approx A$
for the late time behavior.
Then, the luminosity for the late time behavior 
is obtained as
\begin{equation}
	L\approx 
	\frac{1}{48\pi}f_{l}
	\omega_{s}^{1/2}(u_{0}-u)^{-3/2},~ 
	\hat{L}\approx
	\frac{1}{48\pi}f_{l}^2 
	\omega_{s}(u_{0}-u)^{-1}.
	\label{eq:hatl}
\end{equation}
Therefore the luminosity diverges to positive infinity 
for the both fields as $u\to u_{0}$.
The radiated energy is obtained by integrating the luminosity 
with $u$.
The radiated energy for the late time behavior is estimated as
\begin{equation}
E\approx\frac{1}{24\pi}f_{l}\omega_{s}^{1/2}
(u_{0}-u)^{-1/2},~ 
\hat{E}\approx\frac{1}{48\pi}f_{l}^2
\omega_{s}\ln\frac{(M\omega_{s})^{2/3}}{\omega_{s}(u_{0}-u)}.
\end{equation}
Therefore, the amounts of the 
total radiated energy diverge to positive infinity
for the both fields as $u\to u_{0}$.
In realistic situations, we may assume that 
the naked singularity formation 
is prevented by some mechanism and that the quantum particle creation
is ceased at the time $u_{0}-u\approx \Delta t$,
which implies that $G^{\prime\prime}/G^{\prime}$ 
vanishes for $u_{0}-u\alt \Delta t$.
Then, the second term in the 
expression of the luminosity $L$
gives no contribution to the
total radiated energy.
Therefore, the amounts of the total energy for the both fields
are the same, i.e.,
\begin{equation}
E=\hat{E}\approx \frac{1}{48\pi}f_{l}^{2}
\omega_{s}\ln \frac{(M\omega_{s})^{2/3}}{\omega_{s}\Delta t}.
\end{equation}
From the numerical results and physical 
discussion above, we obtain the 
following formula for the late time behavior
\begin{equation}
G(u)\approx A (u-u_{0}) -\frac{4}{3}
A f_{l}\omega_{s}^{1/2}
(u_{0}-u)^{3/2}+\mbox{const}.
\label{eq:GlateLTB}
\end{equation}

For comparison and for a test of our numerical code, 
we have also calculated the function $G$ for the 
Oppenheimer-Snyder (OS) collapse to a black hole
which is given by $F(r)=F_{3}r^3$.
We only display the numerical results for the model with
$F_{3}=1$ and $r_{sf}=0.02$ in an arbitrary unit.
$M$ is given by $M=4\times 10^{-6}$ for this model.
The results are plotted also in Figs. \ref{fg:g_comb} 
and \ref{fg:p_comb}.
Because we cannot define $u_{0}$ for the OS spacetime,
we plot the numerical results by setting $u_{0}=0$. 
For $u>0$, the numerical results show the well-known behavior
\begin{eqnarray}
G(u)&\approx& -\mbox{const}\cdot\exp\left(\frac{-u}{4M}\right)+v_{h},\\
L&\approx&\hat{L}\approx \frac{1}{768\pi M^2}
\label{eq:lHawking},
\end{eqnarray}
where the ingoing null ray with $v_{h}$ is reflected to the outgoing 
null ray which is on the event horizon.
The numerical results of $L$ and $\hat{L}$ for the
OS collapse show a good agreement with Eq.~(\ref{eq:lHawking}).
In Figs.~\ref{fg:g_comb} and \ref{fg:p_comb},
we can find that the assumption concerning the early time and
late time behaviors is justified.

If a back reaction of quantum effects
would not become important until a considerable fraction 
of the total energy of the system is radiated away,
the emitted energy could amount to
$E\sim 10^{54}(M/M_{\odot})\mbox{erg}$.
Thus, the naked singularity explosion would be a new
candidate for a source of ultra high energy cosmic rays.
It may also be a candidate for 
the central engine of a gamma ray burst.
In order to study such possibilities, 
we need to transform the obtained intrinsic quantities
to observed quantities by taking possible reactions
of created energetic particles into consideration.
It is noted that, since we have obtained the formula for 
the late time behavior of the function of $G(u)$,
we can determine the spectrum of the radiation and thereby
estimate the validity of the geometrical optics 
approximation.
We are now obtaining positive evidences for the consistency
~\cite{hin2000}.

We are grateful to H.~Sato for his continuous encouragement.
We are also grateful to T.~Nakamura, H.~Kodama, T.P.~Singh, 
A.~Ishibashi and S.S.~Deshingkar for helpful discussions.
We thank N.~Sugiura for careful reading the manuscript. 
This work was supported by the 
Grant-in-Aid for Scientific Research (Nos. 9204 and 11640273)
and for Creative Basic Research (No. 09NP0801)
from the Japanese Ministry of
Education, Science, Sports and Culture.

\newpage

\begin{figure}
	\centerline{\epsfxsize 8cm \epsfysize 6cm \epsfbox{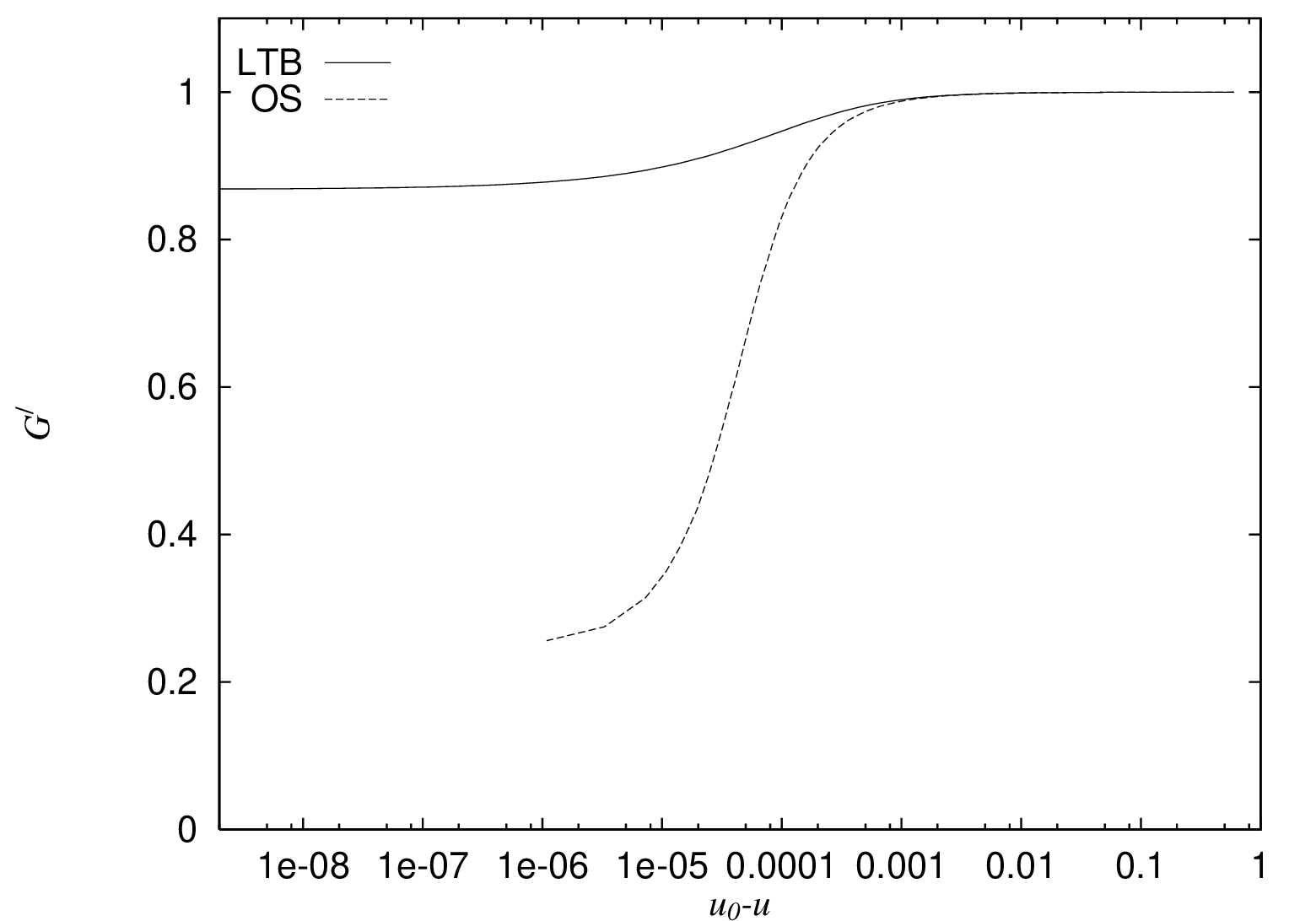}}
~(a)
\end{figure}

\end{document}